\documentclass[numberedappendix,apj]{emulateapj}
\usepackage{amsmath}
\usepackage{rotating}
\slugcomment{Submitted}

\begin{document}
\title{The extent of magnetic fields around galaxies out to $z \sim 1$}

\author{M.L. Bernet\altaffilmark{2,3},F. Miniati\altaffilmark{2}, 
S. J. Lilly\altaffilmark{2}}

\altaffiltext{2}{Physics Department, ETH Zurich, Wolfgang-Pauli-Strasse 27, CH-8093 Zurich, Switzerland}
\altaffiltext{3}{Centre for Astrophysics and Supercomputing, Swinburne University of Technology, Hawthorn, Melbourne, Victoria 3122, Australia}

\email{mbernet@swin.edu.au,fm@phys.ethz.ch,simon.lilly@phys.ethz.ch}

\begin{abstract}
  Radio quasar sightlines with strong MgII absorption lines display
  statistically enhanced Faraday Rotation Measures (RM) indicating the
  presence of additional magneto-active plasma with respect to sightlines
  free of such absorption. In this letter, we use multi-color optical imaging
  to identify the likely galaxies hosting the magneto-active plasma,
  and to constrain the location of the latter with respect to the
  putative parent halo.  We find that all of the sightlines with high
  $|\rm{RM}|$ pass within 50 kpc of a galaxy and that the $|\rm{RM}|$
  distribution for low impact parameters, $D < 50$ kpc, is
  significantly different than for larger impact parameters. In
  addition, we find a decrease in the ratio of the polarization at 21
  cm and 1.5 cm, $p_{21}/p_{1.5}$, towards lower D. These two effects
  are most likely related, strengthen the association of excess
  $|\rm{RM}|$ with intervening galaxies, and suggest that intervening
  galaxies operate as inhomogeneous Faraday screens. These
    results are difficult to reconciliate with only a disk model for
    the magnetic field but are consistent with highly magnetized
    winds associated with MgII systems. We infer strong magnetic
    fields of a few tens of $\mu$G, consistent with values required by
    the lack of evolution of the FIR-radio correlation at high
    redshifts. Finally, these findings lends support to the idea that
  the small scale helicity bottleneck of $\alpha$-$\Omega$ galactic
  dynamos can be significantly alleviated via galatic winds.
\end{abstract}
                                  
\keywords{galaxies: high-redshift --- galaxies: magnetic fields ---
  quasars: absorption lines --- galaxies: evolution}
\footnote{Based on observations made with the ESO Telescopes at the
    La Silla Observatories under programme 082.A-0917 and 085.A-0417.}

\section{Introduction}

The origin and evolution of magnetic fields in galaxies over cosmic
time is observationally still largely unconstrained.  For normal galaxies at
significant look-back times, statistical studies of the Faraday
Rotation effect on luminous polarized background sources provide the
most effective way to probe magnetic fields.  In
\citet{Bernet2008,Bernet2010} we presented evidence that quasars with
strong intervening MgII absorption lines in their optical spectra
(with equivalent width, EW$_0 > 0.3$\AA) have a significantly broader
distribution of Rotation Measure (RM)  than those without.~\citet{Bernet2010} showed that
this was unlikely to be due to any indirect correlation with the quasar
redshift, since the effect was not present for sightlines with weaker
MgII absorption. Since strong MgII absorption is known to be generally associated with the halos of
normal galaxies, the simplest interpretation was that $\sim
10\: \mu$G large scale magnetic fields exist in or around galaxies out
to $z \sim 1.3$ \citep{Bernet2008}.

In this Letter we study the radio properties of background quasars
at different impact parameters from the MgII host galaxies
responsible for the enhanced RM.  A similar approach has been used to study individual nearby galaxies,
e.g. M31 \citep{Han1998}, the LMC \citep{Gaensler2005} and the SMC
\citep{Mao2008}, which have numerous polarized background sources
available. For distant galaxies this is not viable and a
statistical approach is necessary.

Our study reveals that strong magnetic fields are present around
galaxies out to large impact parameters of order 50 kpc. Applying recent
results from the study of MgII systems \citep{Bordoloi2011,Bordoloi2012}, this suggests that the 
ubiquitous winds in high redshift galaxies \citep{Weiner2009,Bordoloi2011,Rubin2010} are highly magnetized.
Furthermore, this finding provides support to the idea that magnetized outflows help
removing small scale helicity from galactic disks~\citep{Shukurov2006},
preventing the quenching of $\alpha$-$\Omega$ dynamo
mechanism~\citep{Vainshtein1992}, thought to generate the large
scale magnetic fields in galactic disks\citep{Brandenburg2005}.

Where required, we assume a concordance cosmology with $h=0.71$,
$\Omega_{M}=0.27$ and $\Omega_{\Lambda}=0.73$.

\section{Observations and data reduction}
\begin{figure*}
\begin{center}
\includegraphics[width=0.9\textwidth]{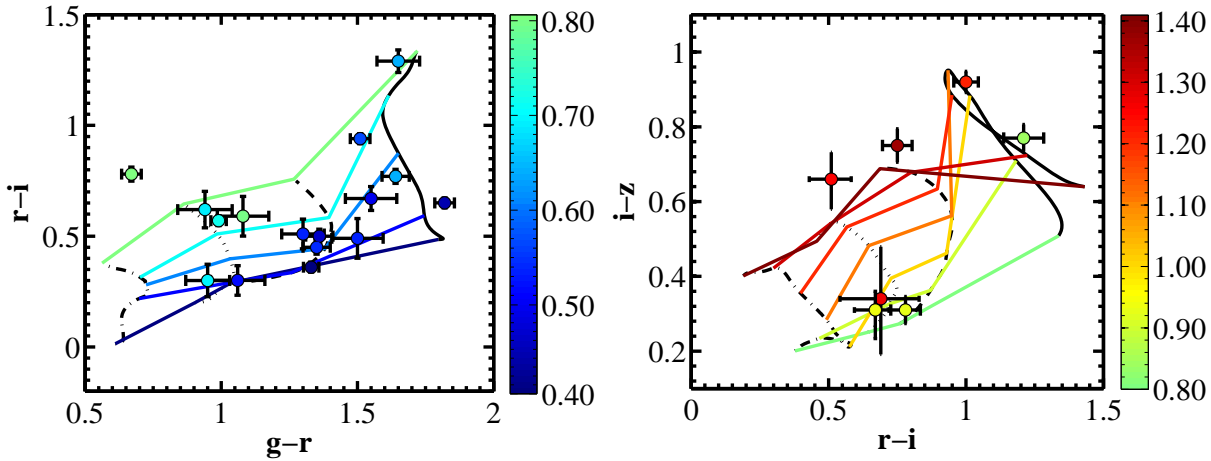}
\caption[Color-color plot of candidate galaxies]{$g-r$
  vs. $r-i$ and $r-i$ vs. $i-z$ plot of identified host galaxies of the MgII
  absorption systems in the redshift range $0.4 < z_{MgII} < 0.8$ and 
  $0.8 < z_{MgII} < 1.4$, respectively. The loci of an elliptical, Sbc, Scd and irregular galaxy are
  shown as solid, dashed, dotted and dashed-dotted lines. The color code indicates the photometric redshift of the galaxies.
  \label{fig:color_color}}
\end{center}
\end{figure*}

We obtained images of the fields of 28 radio quasars with strong
MgII absorption lines selected from the sample of~\citet{Bernet2008}.

Images were taken in three bands with the EFOSC2 instrument at
the NTT in P82 from 30.10.08 - 02.11.08 and in P85 from 15.03.10 -
18.03.10. The quasar fields with absorbers in the
redshift range 0.4 - 0.8 were observed with $g,r,i$ filters and those
with absorbers in the range 0.8 - 1.4 with the $r,i,z$ filters. These
filters were chosen in order to straddle the $4000\: \rm{\AA}$ break
at the redshifts of the absorber. The total exposure times in each
filter varied between 600-6000s depending on the redshift of the
absorbers. We aimed to detect galaxies down to $0.1~\rm{L^{\star}}$ 
\footnote {$L^{\star}$ is the characteristic galaxy luminosity of the 
Schechter function where the power law cuts off.} at
the absorber redshift. To facilitate PSF for subtraction of the quasar,
the camera was used in the 1x1 binning mode giving a pixel scale of
0.12 arcsec per pixel.

In order to identify the MgII host galaxies, we place the galaxies 
in the quasar field on a color-color diagram and compare them
with the theoretical locii of galaxies of different
types computed from spectral energy distributions from \cite{Coleman1980} as a function of redshift.
The photometry of the galaxies was done using Sextractor
\citep{Bertin1996}. The flux was measured in each filter using
circular apertures with diameter of typically 30 pixels, corresponding
to $3.6$ arcsecs. This was reduced when necessary to avoid contamination by neighbouring objects.

To identify the host galaxy we proceed as follows:

\begin{itemize}
\item[i)]Measure the ($r-i$) and ($g-r$) or ($i-z$) colors of all the
  galaxies within 120 kpc of the quasar at the redshift of the absorber. 
  In cases where there was no detection of a galaxy in one
  of the bands, upper limits were calculated for that band.

\item[ii)] All galaxies that have colors in the
  color-color diagram consistent with the locus of SED-types at the redshift
  of the MgII system (see Fig. \ref{fig:color_color}), are considered as candidate host galaxies.
  When in regions of color-color space there is a degeneracy
  between SED-type and redshift a morphological
  classification was done in order to separate these two
  quantities. 

\item[iii)] For objects very close to the quasar, within $\sim 2''$, it
  was not possible to do accurate photometry of the galaxies. We assume that
  all three objects within $\sim 2''$ are the host. (See \cite{Chen2010b}).

\item[iv)] In cases where there were two candidate host galaxies, both
  with consistent colors, the closer one to the quasar was selected when the 
  impact parameter differed by more than a factor two. For cases where they differed
  by less than a factor two the impact parameters were averaged (PKS2204-54, 4C+19.34, 4C+13.46, PKS0506-61, PKS0038-020).

\end{itemize}

The properties of the identified host galaxies are given
in Table \ref{tab:FavGal}.
While there is inevitably a certain arbitrariness to the
identification of the host galaxies, we emphasize that
this process was carried out blind with respect to the RM values of
the quasars in order to preserve the statistical integrity of the
analysis. For the quasar field 4C+06.41
\citep{Lanzetta1995} and 4C+19.44 \citep{Kacprzak2008} spectroscopic redshifts
of galaxies are available and those agree with our choice of host galaxies.

The RMs in this work are selected from the sample of
  \cite{Kronberg2008} at Galactic latitudes $|b| > 30^{\circ}$. At
  least three polarization angles, typically measured at wavelengths
  around 6 cm, were used for the RM determination~\citep[for more details
  see][]{Bernet2012}.

\section{Results}
\label{sec:results}

\subsection{ $|\rm{RM}|$ vs. impact parameters} 

\begin{figure}
\includegraphics[width=0.5\textwidth]{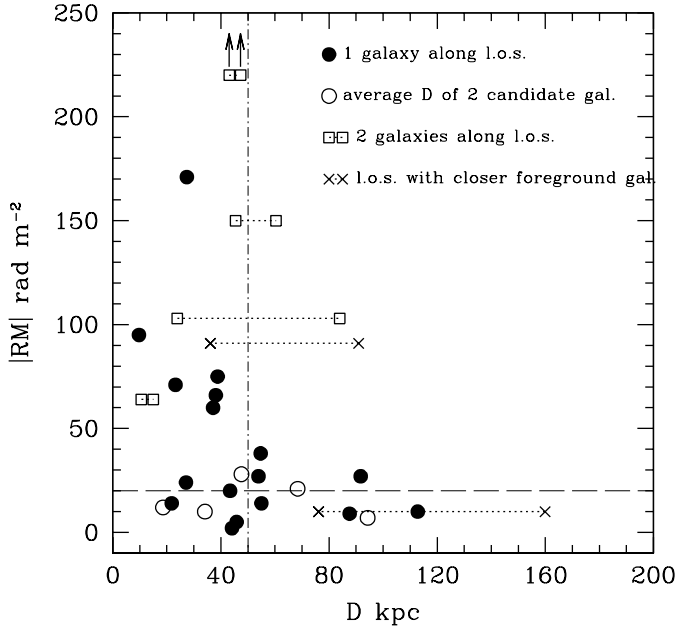}
\caption{Observed $|\rm{RM}|$ vs. impact parameters D of the quasar
  sightlines to the host galaxies of the MgII absorption systems. 
  Solid circles show l.o.s. with one MgII system and with a unique association of
 the galaxy which produces the RM. Open circles show l.o.s. with one MgII system 
but with two possible host galaxies and where the impact parameters differ by less 
than a factor two (shown at average D). The cross symbol corresponds to a l.o.s. 
where next to the galaxy at $z_{MgII}$ a foreground galaxy was detected at smaller 
impact parameter and at $z\ne z_{MgII}$. 
The horizontal dashed line indicates the $20\; \rm{rad}\; \rm{m}^{-2}$ level below 
which Galactic RM dominates. For illustration purposes the $|\rm{RM}|$ value of QSO 
sightline PKS2326-477 is shown at $|\rm{RM}|= 220\; \rm{rad}\; \rm{m}^{-2}$ instead of 
360 $\rm{rad}\; \rm{m}^{-2}$. The typical error on the observed RM is $\sim 3.5\; \rm{rad}\; \rm{m}^{-2}$ (median). 
\label{fig:ObsRM_vs_D}}
\end{figure}

In Fig.~\ref{fig:ObsRM_vs_D} we plot the observed $|\rm{RM}|$, 
  uncorrected for Galactic foreground, for the 28 sightlines
containing strong MgII absorption systems, against the impact
parameters D to the identified host galaxies. Filled circles show
sightlines with a single MgII system and a uniquely identified
intervening galaxy.  Open circles correspond to the case of two
candidate host galaxies plotted at average impact parameter (see
(iv) above). Crossed symbols correspond to the case in which a
foreground galaxy lies closer to the line of sight than the MgII host
galaxy and which is therefore likely to have contaminated the RM
value.  
Open squares indicate quasars sightlines with two strong MgII systems.
Full details about each sightline are provided in Table 1.

Fig.~\ref{fig:ObsRM_vs_D} shows that the $|\rm{RM}|$
increases significantly towards sightlines with smaller impact
parameter. All $|\rm{RM}|$ values $> 50\; \rm{rad}\; \rm{m}^{-2}$ are
at $\rm{D} < 50\; \rm{kpc}$.  At significantly larger impact
parameter, $> 60\; \rm{kpc}$, the RM is mostly given by the Milky Way
contribution and observational error (horizontal dash line), which
amount to $\sim 20\; \rm{rad\; m^{-2}}$ \citep{Bernet2008}. This is similar
to the 68\% percentile spread of the $|\rm{RM}|$ for 
quasars without MgII absorption, i.e. 
to $|\rm{RM}|_{68} \sim 25 \;\rm{rad}\; \rm{m}^{-2}$~\citep{Bernet2010}.
Note that Galactic foreground as, e.g., in \citet{Oppermann2012} does
not introduce any trend with impact parameter, while it contributes
to our sources a median $|\rm{RM}|~ 12.6\; \rm{rad}\; \rm{m}^{-2}$. 

A Pearson' test for correlation for the circles in
Fig. ~\ref{fig:ObsRM_vs_D} shows that $|\rm{RM}|$ and D are
anticorrelated with $\rho=-0.43$, corresponding to a chance
probability of $p=4.8\%$.  A two-sided Kolmogorov-Smirnov (KS) test
for the same data reveals a chance probability that the
$|\rm{RM}|$ distributions below and above 40 kpc 
are drawn from the same distribution of $p=2.2\%$.

This analysis corroborates the above result that the RM contribution
is negligible for sightlines beyond 60 kpc.  We can therefore try and
repeat the KS-test including the two sightlines with closer foreground
galaxies without MgII absorption (x symbols), and also the four
sightlines with two MgII systems (open squares), assuming that only
galaxies with $D<$ 50 kpc contribute RM.  This lowers the above chance
probability to $p=1.5\%$ and 1.9\%, if we split the sample at 40 and
50 kpc, respectively.

Using the sightlines with one MgII absorber at $\rm{D} < 50\; \rm{kpc}$ we
  determine an observed dispersion of RM $\sigma_{obs} \sim 65\;
  \rm{rad}\: \rm{m}^{-2}$.  The contribution from the intrinsic RM, the Milky Way RM, and the 
observational error is found from sightlines without absorbers to be
$\sim 25\; \rm{rad}\: \rm{m}^{-2}$. Subtracting in quadrature
and multiplying by  $(1+z_{MgII})^{2}$, we obtain a rest frame RM dispersion 
for the MgII absorbers $\sigma_{MgII} \sim 150\; \rm{rad}\: \rm{m}^{-2}$.

\subsection{Inhomogeneity of the RM screens}

In \cite{Bernet2012} we studied the effect of depolarisation due to 
inhomogeneous Faraday screens in intervening galaxies at 
redshift $z$, with RM dispersion $\sigma_{RM}$ and covering factor $f_c$.
We predicted a wavelength $\lambda$ dependent effect, 
$p(\lambda^{2})/p_{0}=f_{c}\exp(-2\sigma_{RM}^{2}(1+z)^{-4}\lambda^{4}) +(1-f_{c}),$
where $p_0$ is the intrinsic polarisation.
Since according to Fig.~\ref{fig:ObsRM_vs_D} the observed RM
dispersion increases for smaller impact parameter, we expect the
degree of polarization to follow a similar pattern.

In order to test the depolarization potentially suffered by our
sources, we use the ratio $p_{21}/p_{1.5}$, where $p_{21}$ and
$p_{1.5}$ are the degrees of polarization at 21 and 1.5 cm, from
\cite{Taylor2009} and \cite{Condon1998}, and from \cite{Jackson2010}
and \cite{Murphy2010}, respectively.  Since in general the short and long
wavelength emission originate from different components of the radio
source (i.e. the compact core and the radio lobes), the above ratio is
not a ``measure'' of depolarization.  However, due to the wavelength
dependence of the depolarization effect, statistically the ratio
$p_{21}/p_{1.5}$ will be lower the stronger the depolarizing effect
along the line of sight.

\begin{figure}
\begin{center}
\includegraphics[width=0.5\textwidth]{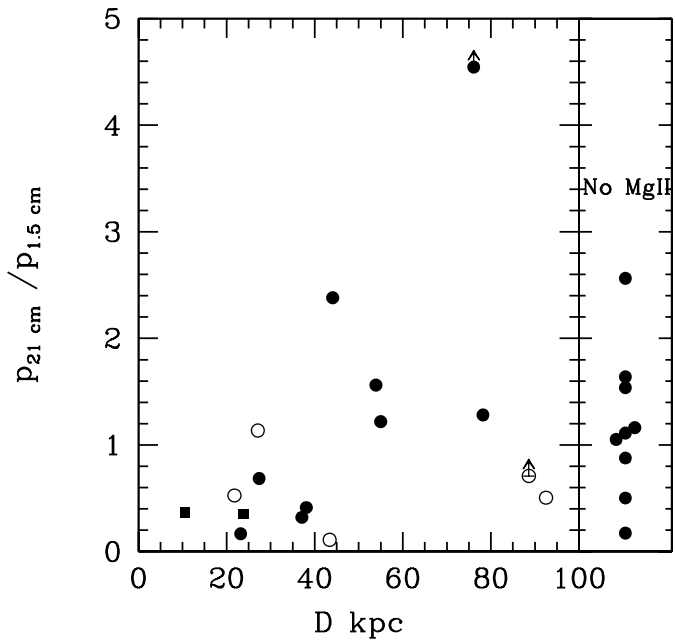}
\caption[Depolarization vs. impact parameters]{Depolarization proxy,
  $p_{21}/p_{1.5}$, as a function of the intervening galaxies' impact
  parameter. Solid circles show QSO at $z > 1.0$ and open circles at 
$z < 1.0$. For the two QSO fields with two intervening galaxies only 
the inner galaxies are plotted and marked as solid squares (see text). 
For comparison the $p_{21}/p_{1.5}$ values of sightlines of QSO at $z > 1$ 
without MgII absorption are plotted in the right hand panel.
\label{fig:DP_vs_D}}
\end{center}
\end{figure}

In Fig.~\ref{fig:DP_vs_D} we plot the ratio $p_{21}/p_{1.5}$ versus
the impact parameters D to the galaxies. Solid/open circles indicate 
if the quasar redshift is $>1.0$ or $<1.0$. Due to the low number of
quasar fields for which both $p_{21}/p_{1.5}$ and D are available,
this plot also includes two sightlines with multiple absorbers, shown as 
solid squares. For one of them (PKS 1143-245) one
galaxy is at 21 kpc and the other at 84 kpc, so it is likely that only
the closer one is contributing. For the other (4C-02.55) both galaxies
are close in, at 10.6 kpc and 11.8 kpc from the sightline,
respectively, so both are likely contributing to the observed RM.

A clear trend is visible in Fig.~\ref{fig:DP_vs_D} in the left hand panel, whereby the lower
the impact parameter, the lower the value of
$p_{21}/p_{1.5}$. A Kendall's $\tau$ test shows that for the overall
sample, $p_{21}/p_{1.5}$ and D are correlated with $\tau=0.30$ and a
chance probability of $11.6\%$. Knowing that quasars with $z_{QSO} < 1.0$ are
intrinsically more depolarized than those above this redshift
\citep{Bernet2012} we split the sample according to this
redshift. This effectively separates out two objects with low
$p_{21}/p_{1.5}$ values at high impact parameters. Repeating now the
test for the $z_{QSO} > 1.0$ sample shows that $p_{21}/p_{1.5}$ and D are
strongly correlated with $\tau=0.56$ and a chance probability of only
1.7 $\%$. For comparison, the $p_{21}/p_{1.5}$ for sightlines from the
whole \cite{Bernet2008} sample that do not have MgII absorption
systems and are at $z_{QSO} >1.0$ are plotted in the right hand panel in Fig.~\ref{fig:DP_vs_D}.

As in~\cite{Bernet2012}, a Faraday screen can be modeled as a
collection of cells of size $l_c$, the magnetic field coherence length.
The (rest frame) RM dispersion of the cells on a Faraday screen
of depth $L$ is then written as:
\begin{equation}
\sigma_{RM}=\sigma_{c}\sqrt{L/l_{c}}\propto Bn_{e}l_{c}\sqrt{L/l_{c}},
\label{Eq:sigmaRM}
\end{equation} 
where $\sigma_{c} \propto Bn_{e}l_{c}$ is the RM of a single cell.  We
can now relate $\sigma_{RM}$ to the (rest-frame) RM dispersion
characterizing sightlines through a galactic Faraday screen, $\sigma_{MgII}$, by
\begin{equation}
\label{eq:RMrelation}
\sigma_{RM}=\sigma_{MgII}\sqrt{N/f_{c}},
\end{equation}
where $N$ is the number of surface cells covering the inhomogeneous
screen and $f_c$ its covering factor.  Using the value of $\sigma_{MgII}$ 
estimated at the end of last section, we see that the decrease in
$p_{21}/p_{1.5}$ towards lower impact parameters can be explained by
the observed increase in $|\rm{RM}|$ alone (Figure
\ref{fig:ObsRM_vs_D}) for $N\sim$ a few. In general, however, the
coherence scale of the RM, i.e. N, and $f_{c}$ might also change as a
function of $D$.

\section{Discussion \& Conclusions}

The simultaneous increase towards small impact parameters of the
  RM dispersion on the one hand, and depolarizing effects as
  probed by $p_{21}/p_{1.5}$, on the other, is consistent with and
  coroborates previous results~\citep{Bernet2008} suggesting the
  presence of significant magnetic fields in and around galaxies at
  distances of several tens of kpc, acting as inhomogeneous Faraday
  screens~\citep{Bernet2012}.

In principle, the observed RM can be attributed to magnetic fields
either in the disk or in the halo of the intervening galaxies. The
RM associated with the disk magnetic field can be probed using a
simple axissymmetric spiral magnetic field model and a distribution of
free electrons, $n_{e}$, both with radial scalelength and vertical
exponential scaleheight of 30kpc and 1.8kpc,
respectively~\citep{Gomez2001,Gaensler2008}, and with normalizations,
$B_{0}=10\: \mu$G, ${n_{e}}_{0}\sim 0.03\: \rm{cm^{-3}}$, at the
galactic center.  A simple Monte-Carlo model for the RM from a sample
of galaxies at random orientation along the line of sight, then shows
that a pure disk magnetic field is not able to account for the
observed $\rm{|RM|}$ vs D reported in Fig.~\ref{fig:ObsRM_vs_D},
unless the normalizations or the vertical/radial scales are made
unrealistically large. This implies that, while the disk magnetic
field may contribute to the observed RM, other components are most likely
present.

It is now established that galactic winds appear in galaxies with star
formation rates at low \citep{Bouche2012}, intermediate
\citep{Bordoloi2011,Weiner2009,Rubin2010} and high redshifts
\citep{Pettini2002}. \cite{Bordoloi2011} finds that MgII absorption in 
foreground edge-on galaxies at $0.5 < z < 0.9$ shows a strong azimuthal
dependence within 50 kpc, indicating the presence of bipolar 
outflows around the disk rotation axis. \citet{Bordoloi2012} shows that
most MgII quasar absorption systems also lie within $45^{\circ}$ of the minor
axis and at $D < 40\: \rm{kpc}$.

Since our MgII systems show statistically the same properties, the
sightlines in our sample will pass through regions above the poles
(this will be soon tested with HST imaging of most of the quasar
fields in our sample).  Further magnetized outflows, although at
significantly smaller distances of a few kpc, have also been detected
in local galaxies \citep{Haverkorn2012}.  Therefore, it is likely that
the high RM values in Fig.~\ref{fig:ObsRM_vs_D} are associated with
magnetized outflows.

Can we get meaningful column densities of free electron $N_{e}$ in the
outflows?  To calculate $N_{e}$ we use a simple wind model based on
the work of~\cite{Bouche2012}.  We assume that the outflow rate is
proportional to the SFR, $\dot{M}_{out}=\eta\, \rm{SFR}$. In addition,
for a biconical geometry and constant outflow velocity $v_{out}$,
$\dot{M}_{out} \simeq \frac{\pi}{2}\mu N_{g}D v_{out} \Theta_{max},$
with $N_{g}$ the gas column density measured at impact parameter D
from a sightline transverse to the outflow, $\Theta_{max}$ the opening
angle and $\mu$ the mean atomic weight~\citep{Bouche2012}.  For a
fully ionized gas, we can then estimate the free electron column
density as, $N_{e} \approx 9\cdot 10^{19} cm^{-2} (\frac{\eta}{0.5} )
( \frac{SFR}{10 M_{\odot} yr^{-1}} ) \times (\frac{v_{out}}{200 km\:
  s^{-1}} )^{-1} (\frac{D}{30 kpc} )^{-1}
(\frac{\Theta_{max}}{30^{\circ}} )^{-1} , $ which is essentially the
same as the obtained estimate in \cite{Bernet2008} for typical MgII
systems, based on HI measurements and an estimate of the ionization
correction.  This shows that the column density at large D can be
substantial in outflows of normal galaxies. Using Eq. \ref{Eq:sigmaRM}
and Eq. \ref{eq:RMrelation} we then infer a magnetic field strength
\begin{align}
B_{\parallel} & = 54 \mu G \bigg (\frac{\sigma_{MgII}}{150 rad\: m^{-2}} \bigg ) \bigg ( \frac{s}{10 kpc} \bigg ) \bigg (\frac{L}{10 kpc} \bigg )^{0.5} \nonumber \\
& \times \bigg (\frac{f_{c}}{0.5} \bigg )^{-0.5} \bigg ( \frac{N_{e}}{9 \cdot 10^{19}{cm}^{-2}} \bigg )^{-1} \bigg ( \frac{l_{c}}{3 kpc}  \bigg )^{-1.5},\label{Best:eq}
\end{align}
where we have set the number of independent RM cells to $N=s^{2}/l_{c}^{2}$,
with $s$ the projected linear size of the source at $z_{MgII}$ and
have assumed that the scale of the RM fluctuation, $l_{c}$, is the same
along the sightline and in the plane of the sky. This value of $B$ is higher
than the $\sim 10\:\mu$G obtained in \cite{Bernet2008} because here we
consider that the RM screens are inhomogeneous. 

This simple analysis shows that magnetized winds can account for the
observed $\rm{|RM|}$ at large impact parameter reported in
Fig.~\ref{fig:ObsRM_vs_D}. This requires a magnetic field strength $B$
of several tens of $\mu$G, which is considerably larger than the few
$\mu$G fields observed in large scale outflows in nearby galaxies,
e.g. NGC5775 \citep{Tuellmann2000}, NGC4666 \citep{Dahlem1997}. 

 However, fields of such strength are required in high redshift
  galaxies by the lack of evolution in the FIR-Radio
  correlation~\citep{Condon1992,Ivison2010,Sargent2010}.  In fact, the
  radio emitting electrons would otherwise mostly radiate their energy
  through inverse Compton scattering on the CMB and/or starlight, both
  significantly higher at $z\sim 1$ due to cosmological expansion and
  a $\sim 10\times$ higher SFR, respectively.  Since this is not
  observed, the total magnetic field must have been larger by $\propto
  \max[{\rm SFR}^{1/2},(1+z)^2]\sim 4$. If this was the case for both
  large and small scale fields, then values of order of a few tens of
  $\mu$G are found for $z\sim 1$ galaxies, if our reference low-z
  field is the Galactic one at several $\mu$G.

The existence of magnetized outflows in normal galaxies has important
implications for $\alpha$-$\Omega$ models of galactic
dynamos. Galactic dynamos are thought to be responsible for the origin
of large scale magnetic fields in spiral galaxies.  However, their
efficiency can be severely limited by conservation of magnetic
helicity once the small scale magnetic field reaches equipartition
with the small scale kinetic energy of interstellar
gas~\citep{Vainshtein1992,Brandenburg2005}. However, galactic winds
can transport magnetic helicity away from the plane of the galaxy and
restore the efficiency of $\alpha$-$\Omega$ dynamos, as proposed
by~\citet{Shukurov2006}. Assuming again coexistence of large and small
scale fields, our observations support the occurrence of
this process and could represent a first observational link between
galactic dynamos and magnetized winds at intermediate redshifts.

Understanding the nature of magnetic fields in and around intermediate redshift
galaxies deserves further work.

\section{Acknowledgements}
We thank an anonymous referee for his valuable comments. 
This research has been supported by the Swiss National Science
Foundation and made use of observational facilities of the European
Southern Observatory (ESO).

\begin{sidewaystable}[ht]
\begin{scriptsize}
    \centering
    \caption[Properties of parent galaxies]{Properties of candidate host galaxies}
       \begin{tabular}{lccccccccccccccc}
        \hline
        \hline
        & & & & & & & & & & & & & & & \\
 quasar &$z_{quasar}$ &  RA & Dec &$z_{MgII}$ & $W_{0}(2796)$   & $\Delta \rm{RA}$ & $\Delta \rm{Dec}$        & Ang. sep. & D &$m_{g}$&$m_{r}$&$m_{i}$&$m_{z}$&  type & symbol \\

     &         & (J2000)& (J2000)&     & (\AA)              & arcsec  &  arcsec        & arcsec  & kpc   &      &       &       &      &    & \\
        (1) & (2) &(3) & (4) & (5)& (6)                     &(7) & (8) & (9)  & (10) & (11) & (12) & (13) & (14) & (15) & (16)        \\
       & & & & & & & & & & & & & & &  \\
        \hline
      & & & & & & & & & & & & & & &   \\
4C-02.55 &1.043 &12:32:00.0 & -02:24:05   & 0.39524 & 2.03  &0.6 &-1.9      &2.0  & 10.6     & ...   & ...  & ...  & ...                & ucl. & $\Box$   \\
4C-02.55 &...   &...        & ...         & 0.75689 & 0.30   &1.6 &-0.1     &1.6  & 11.8     & ...   & ...  & ...  & ...                & ucl. & $\Box$   \\
MRC0122-003 & 1.07& 01:25:28.8& -00:05:56 & 0.39943 & 0.47  & -16.8 & 24.8    & 30.0   & $159.9^{a)}$  & ...  & ...  & ... & ...        & ucl. &  x        \\
MRC0122-003 & ... &  ...      & ...       & 0.3971  & ...   & -8.4  & -12.1   & 14.7  & $76.1^{b)}$    &22.21  &20.88  & 20.52 & ...    & Sbc  &  x        \\ 
PKS2326-477 & 1.299& 23:29:17.7&-47:30:19 & 0.43195 & 0.38  & 3.7   & -6.8    &7.7  & 43.1   & ...     &21.49 &20.92 &20.64             & Ell &  $\bullet$ \\
PKS2326-477 & ... & ... & ...             &1.26074 & 0.66   & 1.6   &  5.4    &5.6  & 47.3   & ...     &24.83 &24.14 &23.80             & Sbc &  $\Box$   \\
4C+06.41 &1.270 &10:41:17.1 &+06:10:17    &0.44151 & 0.69   & 9.4   &  2.2    &9.7  & 55.0   & 22.96 &21.14 &20.49 & ...                & Ell &  $\Box$   \\
4C+19.44 &0.720 &13:57:04.4 &+19:19:07    & 0.45653 & 0.85  & 1.6   &  7.8    &7.9  & 45.8   & 23.03 &21.67 &21.17 & ...                & Sbc &  $\bullet$\\
PKS1244-255 &0.633 &12:46:46.8 &-25:47:49 &0.49286 & 0.68   & -3.9  &  2.3    &4.5     & 27.1   & 24.47 &22.92 &22.25 &                 & Sbc &  $\bullet$\\
OC-65 &0.733 & 01:41:25.8 & -09:28:44     &0.50046 & 0.53   & -0.4  & -1.4    &1.5  &  9.1   & ...   & ...  & ...  & ...                & ucl.&  $\bullet$\\  
PKS0130-17 & 1.022& 01:32:43.5 &-16:54:49 &0.50817 & 0.59   & -3.9  & -4.7    &6.0  & 37.1   & 23.50 &22.44 &22.14 & ...                & Scd &  $\bullet$\\
4C+19.34 &0.828 &10:24:44.8 & +19:12:20   &0.52766 & 1.00   & -5.6  & 4.0     &6.9  & 43.2   & 24.83 &23.33 &22.84 & ...                & Sbc &  $\circ$\\
PKS1615+029 &1.339 &16:17:49.9 &02:46:43  & 0.52827 & 0.31  & 5.7   & -2.3     &6.2  & 38.8   & 24.31 &23.01 &22.50 &...                & Sbc &  $\bullet$\\
4C+01.24  &1.024 &09:09:10.1 &+01:21:36   & 0.53587 & 0.44  & -6.8  & 0.9     &7.0  & 44.1   & 23.06 &21.71 &21.26 & ...                & Sbc &  $\bullet$\\
PKSB1419-272&0.985 & 14:22:49.2&-27:27:56 & 0.55821 & 0.44  & 11.2  & 8.0     &13.7 & 88.6   & 23.36 &21.85 &20.91 & ...                & Ell &  $\bullet$\\
OX+57 & 1.932 & 21:36:38.6 & +00:41:54    & 0.62855 & 0.60  & 5.6   & -0.3   & 5.6  & 38.1   & ...   & ...  & ...  & ...                & ucl.&  $\bullet$\\
OX-192 & 0.672 &21:58:06.3 &-15:01:09     & 0.63205 & 1.40  & 2.8   & 1.6    &3.2  & 21.8   & 24.33 &22.68 &21.39 &  ...                & Ell &  $\bullet$\\
PKS0420-01 & 0.915& 04:23:15.8 &-01:20:33 & 0.63291 & 0.77  & 5.7   & -13.3   &13.5 & 92.5   & 24.09 &22.45 &21.68 & ...                & Ell &  $\bullet$\\
3C208 &1.112 &08:53:08.6  &+13:52:55      & 0.65262 & 0.62  & 0.3   & -6.6    &6.6  & 45.5   & ...   &21.90 &20.70 &20.34               & Ell &  $\Box$ \\
3C208 & ...  & ...        & ...           & 0.93537 & 0.40  & -3.5  & 6.7     &7.6  & 60.4   & ...   &23.07 &22.40 &22.09               & Scd &  $\Box$ \\
PKS0038-020 &1.178&00:40:57.6 & -01:46:32 & 0.68271 & 0.35  & -10.8 & -3.8    &11.5  & 78.2  & 23.82 &22.88 &22.26 & ...                & Scd &  $\circ$ \\
PKS2204-54 & 1.206 &22:07:43.7 &-53:46:34 & 0.6877 & 0.73   & -9.7  & -5.3    &11.1 & 78.7   & 23.35 &22.40 &22.10 & ...                & Scd &  $\circ$ \\
PKS0839+18 &1.270 &08:42:05.1 &+18:35:41  &0.71118 & 0.56   & -5.4  &  5.4    &7.6  & 54.7   & 22.70 &21.71 &21.14 & ...                & Scd &  $\bullet$\\
4C+13.46 &1.139 &12:13:32.1 & 13:07:21    & 0.77189 & 1.29  & -0.6   & 1.8    &1.9  & 14.1   & ...   & ...  & ...  & ...                & ucl.&  $\circ$\\
4C+6.69 & 0.99 &21:48:05.4 & +06:57:39    & 0.79086 & 0.55  & 0.3 & -5.8    &5.8  & 43.4   & 23.82 &22.74 &22.15 & ...                  & Scd &  $\bullet$\\
PKS0402-362 & 1.417& 04:03:53.7&-36:05:02 & 0.79688 & 1.80  & 1.8 &  2.5    &3.1  & 23.2   & 23.28 &22.61 &21.83 & ...                  & Scd &  $\bullet$\\
PKS2223-05 &1.404 & 22:25:47.2 &-04:57:01 &0.84652 & 0.60   & -2.1 &  6.7    &7.0  & 53.9   & ...      &23.47 &22.26 &21.49             & Ell &  $\bullet$\\
PKS0506-61 & 1.093 &05:06:43.9 &-61:09:41 & 0.92269 & 0.49  &1.8 & 4.0    &4.4  & 34.6   & ...      &23.42 &22.64 &22.33                & Scd &  $\circ$\\
PKS0112-017&1.365 &01:15:17.1 &-01:27:05  & 1.18965 & 0.90  &2.0 &-2.6    &3.3  & 27.4   & ...      & ...  & ...  & ...                 & ucl.&  $\bullet$\\
PKS0332-403 & 1.445& 03:34:13.7&-40:08:25 & 1.20898 & 0.79  & 3.4 & 10.3    &10.9 & 91.0   & ...    &23.66 &22.66 &21.74                & Ell &  x\\
PKS0332-403 & 1.445& 03:34:13.7&-40:08:25 & $~0.8$ &   & -3.4 & 3.3    & 4.8 & $36.1^{c)}$   & ...   &23.31 &21.93 &21.43               & Ell &  x\\  
PKS1143-245 &1.940 & 11:46:08.1&-24:47:33 & 1.24514 & 0.30  &10.0 &0.3     &10.0 & 84.0   & ...      &23.70 &23.19 &22.53               & Sbc &  $\Box$\\
PKS1143-245 & ... & ... &  ...            & 1.52066 &0.46    & 2.2 & -1.7     &2.8  & 23.9   & ...   & ...  & ...  & ...                & ucl.&  $\Box$\\
OQ135 &1.612&14:23:30.1&+11:59:51         & 1.36063 & 0.51  & -12.9 & -3.1   &13.3 &112.8   &  ...   &23.62 &22.87 &22.12               & Sbc &  $\bullet$\\
 \hline
quasar fields with alternative candidate galaxies  & & & & & & & & & & & & & & &  \\
 \hline

PKS0130-17     & 1.022 & 01:32:43.5 &-16:54:49    & 0.50817 & 0.59    & 11.8 & 6.0   & 13.2  & 81.0      &22.61     &21.76  &21.42     &  ...    &   Sbc & ...\\
4C+19.34       &0.828  &10:24:44.8 & +19:12:20    & 0.52766 & 1.00  & 2.2  & -3.4  & 4.0  & 25.0      &  ...        &  ...  & ...      &  ...    &   ucl.& ... \\
PKS1615+029    &1.339  &16:17:49.9 &02:46:43      & 0.52827 & 0.31    & 3.9  &  4.8  &  6.2  & 38.8       & 22.66    & 21.19 &20.43     & ...    &   Ell & ...\\
4C+01.24       &1.024  &09:09:10.1 &+01:21:36     & 0.53587 & 0.44  & 11.7 & 10.3  & 15.5  & 97.8      &24.16     &22.53  &21.80     &   ...     &   Ell & ...\\
PKS0038-020    & 1.178 &00:40:57.6 & -01:46:32    & 0.68271 & 0.35     & 13.2 &8.3   & 15.6  & 110.3       &23.43     &22.64  &22.06     & ...   &  Scd  & ...\\
PKS2204-54     & 1.206 &22:07:43.7 &-53:46:34     & 0.6877  & 0.73     &3.5  &-7.4   &  8.2  & 58.1       &24.12     &24.04  &23.56     &  ...   &  Irr  & ...\\
PKS0839+18     &1.270  &08:42:05.1 &+18:35:41     & 0.71118 & 0.56   &-3.3 &7.5   &  8.1  & 58.2       &24.03     &22.64  &21.91     &  ...      &  Sbc  & ...\\
4C+13.46       &1.139  &12:13:32.1 & 13:07:21     & 0.77189 & 1.29 &3.1 & 0.4   &  3.1  & 23.0       &24.13     &22.30  &21.14     &   ...       &  Ell  & ...\\
PKS0506-61     & 1.093 &05:06:43.9 &-61:09:41     & 0.92269 & 0.49    &-7.6 & -1.4  &  7.7  & 60.6       & ...         &23.08  &22.39     &22.07 &  Scd  & ...\\
PKS1143-245    &1.940  & 11:46:08.1 &-24:47:33    & 1.52066 & 0.46    &-9.1 & -3.6   &  9.7  & 82.9     &  ...         &23.61  &23.42     &22.79 &  Scd  & ...\\

\hline
  & & & & & & & & & & & & & & & \\
  & & & & & & & & & & & & & & & \\
 & & & & & & & & & & & & & & & \\
 & & & & & & & & & & & & & & & \\
 & & & & & & & & & & & & & & & \\
  & & & & & & & & & & & & & & &  \\
 & & & & & & & & & & & & & & & \\
 & & & & & & & & & & & & & & & \\
& & & & & & & & & & & & &  & &\\
 & & & & & & & & & & & & & & & \\
 & & & & & & & & & & & & & & & \\
  & & & & & & & & & & & & & & &  \\
 & & & & & & & & & & & & &  & &\\
 & & & & & & & & & & & & &  & & \\
& & & & & & & & & & & & &  & & \\
 & & & & & & & & & & & & & & & \\
 & & & & & & & & & & & & &  & &\\
  & & & & & & & & & & & & & & & \\
 & & & & & & & & & & & & &  & &\\

\end{tabular}

    \label{tab:FavGal}
\begin{flushleft}
COLUMNS. - (1) Name of the source, (2) Redshift of the source, (3),(4) quasar's coordinates, (5) Redshift of the MgII system, (6) Equivalent width $W_{0}(2796)$,(7),(8) Coordinates of galaxies with respect to quasar (9) Angular separation, (10) Impact parameter D, (11),(12),(13),(14) Apparent magnitudes in g,r,i,z, (15) Type of galaxy, (16) Symbol used in Fig. \ref{fig:ObsRM_vs_D}.  
$^{a)}$ from \cite{Chen2001} 
$^{b)}$ weak MgII absorber at $z=0.3971$
$^{c)}$ galaxy at $z~0.8$ 
\end{flushleft}
\end{scriptsize}
\end{sidewaystable}

\bibliographystyle{plainnat}

\end{document}